# In-Circuit Differential-Mode Impedance Extraction at the AC Input of a Motor Drive System

Arjuna Weerasinghe, Zhenyu Zhao, Fei Fan, Pengfei Tu, and Kye Yak See
School of Electrical and Electronic Engineering
Nanyang Technological Univeristy
Singapore
harshaya001@e.ntu.edu.sg

*Abstract*—The in-circuit differential-mode (DM) impedance at the AC input of a motor drive system (MDS) serves as a key parameter to evaluate and estimate the DM electromagnetic interference (EMI) noise caused by the switching of power semiconductor devices in the MDS. This paper discusses a single-probe setup (SPS) with frequency-domain measurement to extract the in-circuit DM impedance of an MDS under its different operating modes. The advantages of the SPS are its non-contact measurement and simple structure.

## I. Introduction

Power switching devices in a motor drive system (MDS) generally produce conducted electromagnetic interference (EMI) noise. The EMI noise will propagate from the AC input of the MDS to the grid, thereby affecting the normal operation of other grid-connected electrical assets [1]. To evaluate and estimate the EMI noise, the differential-mode (DM) and common-mode (CM) equivalent noise models of the MDS can be constructed [2], [3]. These noise models are usually represented by the respective DM and CM equivalent noise sources with internal impedances; thus, these internal impedances need to be extracted. Compared to off-circuit impedance measurement, in-circuit impedance measurement brings more realistic results for noise evaluation and estimation. This is because the impedances of the MDS under its actual operating modes deviate significantly from those when the MDS is off-circuit [4].

For the in-circuit impedance measurement, there are mainly three approaches, namely the voltage-current (V-I) measurement [5]-[7], the capacitive coupling [8]-[10], and the inductive coupling [11]-[13]. The V-I measurement approach measures voltage and current using respective sensors and taking Ohm's Law as the definition, the online impedance of an energized electrical system under test (SUT) is calculated [5]. In this approach, the harmonics that already present in the SUT can be used as the test signal [6], or an externally injected signal can be used [7]. The capacitive coupling approach extracts the in-circuit impedance of an energized SUT by using coupling capacitors with an impedance analyzer (IA) [8] or a vector network analyzer (VNA) [9]. The coupling capacitors are connected between the IA/VNA and the energized SUT to provide a low-impedance path for the high-frequency test signal, but block DC or power frequencies [10]. The inductive coupling approach usually adopts two clamp-on inductive probes, a signal generator, and a measuring instrument. One of the inductive probes is used to inject an excitation test signal, and the other inductive probe is used to receive the response from the SUT for the given test signal. By establishing the electrical relationship between the excitation and response signals, the in-circuit impedance of the energized SUT can be obtained [11].

Among the three methods, the voltage sensor used in the V-I measurement approach and the coupling capacitors used in the capacitive coupling approach need to be in direct electrical contact with the energized SUT, which necessitates special provision for such contact, especially when the SUT is powered by high-voltages. In contrast, the measurement setup of the inductive coupling approach does not have direct electrical contact with the energized SUT. Therefore, this approach simplifies the on-site implementation without producing electrical safety hazards.

The inductive coupling approach was first proposed for in-circuit impedance measurement of power lines [11]. Subsequent improvements extended its usage to many other applications [3], [4], [12]. The classic measurement setup of this approach (named "two-probe setup") consists of two clamp-on inductive probes and a frequency-domain measurement instrument, such as a VNA with sweep frequency excitation [12]. Recent developments proposed time-varying in-circuit impedance monitoring using a time-domain measurement instrument in the two-probe setup (TPS) [13]. However, it can only monitor the in-circuit impedance at one frequency at a time. Compared with the frequency-domain instrumentation, it is rather laborious to measure the in-circuit impedance in a wide frequency range. Regardless of the frequency-domain and the time-domain measurements, the two inductive probes will produce direct probe-to-probe coupling, which contaminates the final measurement. Although a calibration technique has been proposed to de-embed this impact, it still cannot eliminate this coupling fundamentally [14]. To fundamentally eliminate the probe-to-probe coupling, a single-probe setup (SPS) with frequency-domain measurement has been developed recently [15], which is basically composed of one clamp-on inductive probe and a VNA. This setup can also include a signal amplification and protection (SAP) module to enhance its signal-to-noise ratio (SNR) and ruggedness, making it a good candidate even for challenging high power environments with significant background noise (e.g. harmonics) and power surges. In view of the above-mentioned merits, this paper discusses the practical application of the SPS to extract the in-circuit DM impedance at the AC input of an MDS under the MDS's different operating modes.

The organization of this paper is as follows. Section II introduces the details of using the SPS to extract the in-circuit DM impedance at the AC input of the MDS. Using a commercially available MDS as a test case, Section III shows the measured in-circuit DM impedances of the MDS under its different operating modes. Finally, Section IV summarizes this paper.

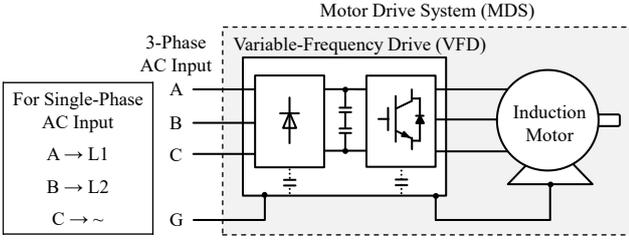

Fig. 1. Schematic diagram of a typical MDS.

## II. IN-CIRCUIT DM IMPEDANCE EXTRACTION AT THE AC INPUT OF AN MDS

As shown in Fig. 1, a typical MDS is composed of a variable frequency drive (VFD) and a three-phase induction motor with cables in between, in which the AC input of the MDS is usually either 3-phase or single-phase depending on the application scale. To extract the in-circuit DM impedance at the AC input of the MDS ($Z_{DM, MDS}$) through the SPS, the experimental setup shown in Fig. 2 is applied. The MDS is connected to the AC power via a line impedance stabilization network (LISN), in which the LISN is to provide a stable and well-defined impedance at the AC power side [4] and to prevent the test signal leaking into the power grid and affecting the operation of grid connected sensitive electrical devices. The ground connection between the MDS and the LISN is left open. The switching of power semiconductor devices in the VFD produces DM noise at the AC input of the MDS, which leads to a DM current path formed by the MDS, power cables, and LISN. From Fig. 2, the SPS includes a VNA, a clamp-on inductive probe, and an SAP module. The SAP module comprises of a signal amplifier, a directional coupler, a surge protector and two attenuators (ATT 1 and ATT 2). The SAP module is used because the MDS usually suffers significant background noise (e.g. harmonics) and power surges. For measuring the in-circuit $Z_{DM,MDS}$, the inductive probe is clamped on one of the power cables with the position marked as $c$-$c'$.

The DM equivalent circuit of Fig. 2 is shown in Fig. 3, in which $V_{DM,MDS}$ represents the equivalent DM noise voltage source of the MDS; $Z_{DM,CABLE}$ represents the equivalent DM loop impedance formed by the power cables; $Z_{DM,LISN}$ represents the equivalent DM impedance of the LISN. To extract the in-circuit $Z_{DM,MDS}$, the signal source of the VNA generates a sweep-frequency test signal, which is amplified by the signal amplifier and then injected into the DM path via the inductive probe. Using the direction coupler, the incident wave and the reflected wave of the test signal can be separated and measured by the two receivers of the VNA, respectively. The two attenuators are used to ensure that the power of the measured test signal is within the permissible range of the receivers. The surge protector is added to protect the VNA from power surges in the MDS and from line events.

Based on the network analysis theory, Fig. 4 shows the cascaded two-port network of Fig. 3 from $m$-$m'$ [15]. $\varGamma_m$ is the reflection coefficient observed at $m$-$m'$, which is obtained directly from the VNA by the incident and the reflected waves of the test signal measured by the two receivers [16]. $N_{IP}$ represents the two-port network of the inductive probe, where $L_{lk}$ and $C_p$ denote its leakage inductance and equivalent parasitic capacitance, respectively. $N_{LISN-CABLE}$ represents the two-port network formed by the LISN and power cables. Since $N_{IP}$ and $N_{LISN-CABLE}$ are cascaded, the resulting two-port

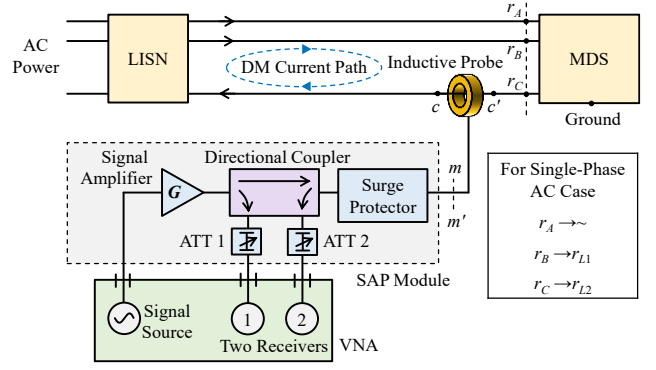

Fig. 2. Experimental setup to extract the in-circuit DM impedance at the AC input of the MDS through the SPS.

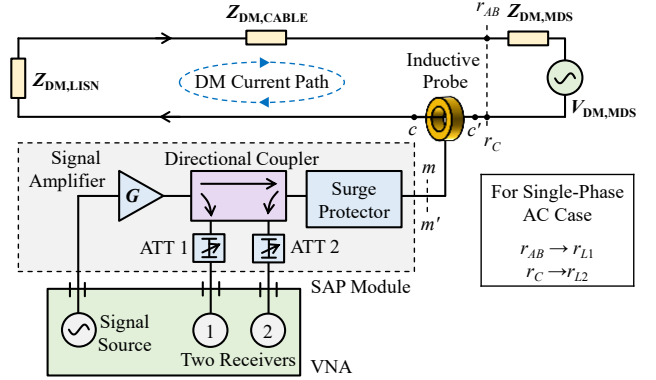

Fig. 3. DM equivalent circuit of Fig. 2.

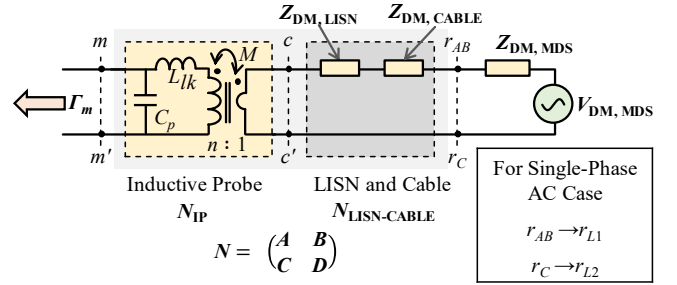

Fig. 4. Cascaded two-port networks representation of Fig. 3.

network $N$ can be expressed in terms of transmission (ABCD) parameters as:

$$N = N_{IP} \cdot N_{LISN-CABLE} \quad (1)$$

The relationship between $Z_{DM,MDS}$ and $\varGamma_m$ can be established according to the ABCD parameters of $N$ as follows:

$$Z_{DM,MDS} = \frac{k_1 \cdot \varGamma_m + k_2}{\varGamma_m + k_3} \quad (2)$$

where

$$k_1 = -\frac{Z_0 \cdot D + B}{Z_0 \cdot C + A} \quad (3)$$

$$k_2 = -\frac{Z_0 \cdot D - B}{Z_0 \cdot C + A} \quad (4)$$

$$k_3 = \frac{Z_0 \cdot C - A}{Z_0 \cdot C + A} \quad (5)$$

From (2), if $k_1$, $k_2$ and $k_3$ are known, $Z_{DM,MDS}$ can be determined by the extracted $\varGamma_m$ from the VNA. As observed in (3)-(5), $k_1$, $k_2$ and $k_3$ are frequency-dependant parameters determined by the *ABCD* parameters of $N$ and $Z_0$, which are unique for a given setup of SPS, LISN, and power cables. To extract $k_1$, $k_2$ and $k_3$, a pre-measurement characterization procedure has been well reported in [15] and only a brief description is given here. To perform the pre-measurement characterization, the open, short and 50Ω resistive load (OSL) conditions are respectively realized at the position $r_{AB}$-$r_C$. The characterization is performed when the AC power is "off" because the LISN's DM impedance and the cables' DM impedance remain unchanged regardless of the "on" or "off" condition of AC power. Based on the characterization, $k_1$, $k_2$ and $k_3$ can be finally determined as follows:

$$k_1 = 50 \cdot \frac{\Gamma_L - \Gamma_O}{\Gamma_L - \Gamma_S} \quad (6)$$

$$k_2 = 50 \cdot \Gamma_S \cdot \frac{\Gamma_O - \Gamma_L}{\Gamma_L - \Gamma_S} \quad (7)$$

$$k_3 = -\Gamma_O \quad (8)$$

where $\Gamma_O$, $\Gamma_S$, and $\Gamma_L$ denote the measured reflection coefficients by the VNA respectively, during the open ($Z_{AB-C} \to \infty$), short ($Z_{AB-C} = 0$), and 50 Ω load ($Z_{AB-C} = 50\angle 0°$ Ω) conditions at $r_{AB}$-$r_C$.

TABLE I
SPECIFICATIONS OF THE MDS, LISN, PROBE, VNA, AND SAP

| Instrument | Specifications |
| --- | --- |
| VFD | TECO L510s (No built-in EMI filter) |
| Induction Motor | RMS8024/B3 (4 pole, 3 Phase, 0.75 kW, 50 Hz) |
| Cables | VFD to Motor: 60 cm<br>LISN to VFD: 100 cm |
| LISN | Electro-Metrics MIL 5-25/2 (100 kHz-65 MHz) |
| Inductive Probe | SOLAR 9144-1N (4 kHz – 100 MHz) |
| VNA | Omicron Bode 100 (1 Hz-40 MHz) |
| Signal Amplifier | Mini circuits LZY–22+ (100 kHz – 200 MHz) |
| Directional Coupler | DC3010A (10 kHz – 1 GHz) |
| Surge Protector | SSC-N230/01 |
| Attenuator 1 | AIM-Cambridge 27-9300-6 (6 dB) |
| Attenuator 2 | AIM-Cambridge 27-9300-3 (3 dB) |

TABLE II
OPERATING MODES OF THE MDS

| Mode | Control Mode | Speed |
| --- | --- | --- |
| Mode 1 | V/F | 10 Hz |
| Mode 2 | V/F | 30 Hz |
| Mode 3 | V/F | 50 Hz |
| Mode 4 | SLV | 10 Hz |
| Mode 5 | SLV | 30 Hz |
| Mode 6 | SLV | 50 Hz |

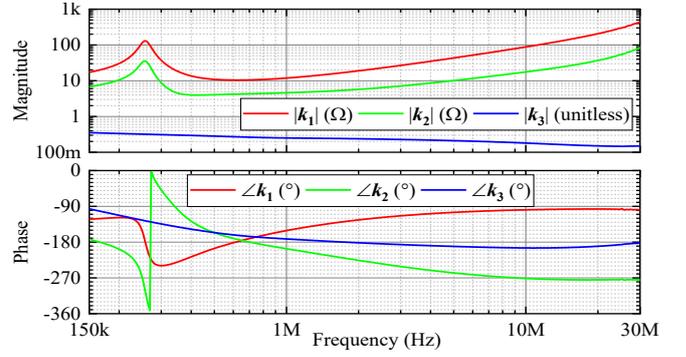

Fig. 5. Frequency-dependant parameters for the selected setup.

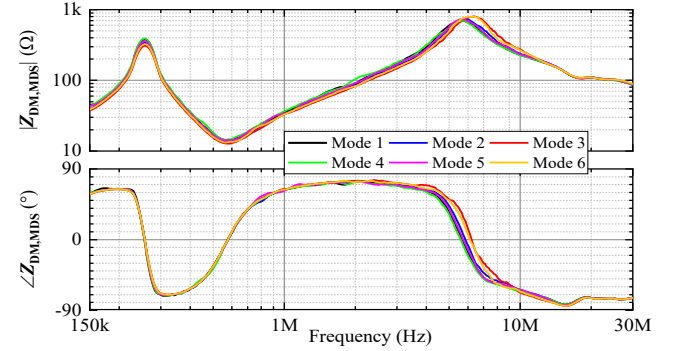

Fig. 6. $Z_{DM,MDS}$ measured under various operating modes of the MDS.

### III. EXPERIMENTAL RESULTS

The test setup shown in Fig. 2 is realized with the equipment shown in Table I to measure the in-circuit DM impedance of an MDS with a single-phase AC input connected through a LISN to the 230-V 50-Hz power grid. The OSL pre-measurement characterization step is carried out without connecting the MDS at $r_{L1}$-$r_{L2}$. The frequency-dependent $k_1$, $k_2$ and $k_3$ for the selected setup are calculated according to (6)-(8) and are visualized in Fig. 5 in the frequency range from 150 kHz to 30 MHz. After that, the actual in-circuit DM impedance ($Z_{DM,MDS}$) measurement can be performed for the MDS by (2).

$Z_{DM,MDS}$ is obtained under various operating modes of the MDS to study the respective correlation. The chosen VFD supports voltage/frequency (V/F) control mode and sensorless-vector (SLV) control mode. Under each VFD control mode the speed (output frequency) setting is changed from 10 Hz to 50 Hz as defined in Table II. The respective measurement results are presented in Fig. 6. For better clarity, the results are presented in Fig. 7- Fig. 11. The obvious from Fig. 7- Fig. 9 is that the VFD control mode does not influence $Z_{DM,MDS}$ for the entire frequency range from 150 kHz to 30 MHz. In contrast, Fig. 10 and Fig. 11 reveal that the speed setting affects $Z_{DM,MDS}$ at certain frequencies ranges marked by grey box indicating the inconsistency. However, the observed impedance change is rather small.

### IV. CONCLUTION

This study covers the practical application of the SPS to extract the in-circuit DM impedance of an MDS under six operating modes encompassing V/F and SLV control modes at three speed settings (10 Hz, 30 Hz, and 50 Hz). A commercially available MDS has been tested and the in-circuit DM impedance measurements from 150 kHz to 30 MHz have been presented. According to the experimental results, it can be concluded that the V/F control mode and SLV control mode does not influence the in-circuit DM impedance of an MDS at AC input, but the speed setting does have a certain impact in some frequency ranges. Reasons for above conclusion will be analytically studied in future.

### ACKNOWLEDGMENT

This research work was conducted in the SMRT-NTU Smart Urban Rail Corporate Laboratory with funding support from the National Research Foundation (NRF), SMRT and Nanyang Technological University; under the Corp Lab@University Scheme.

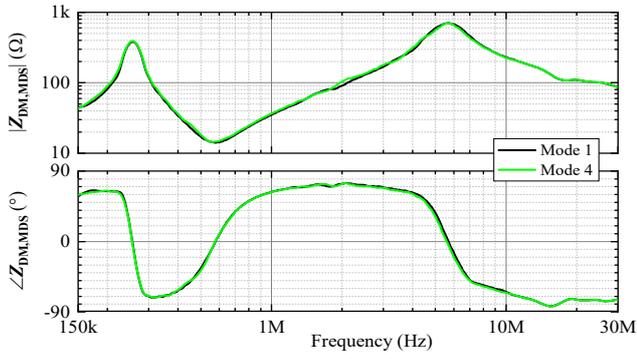

Fig. 7. $Z_{DM,MDS}$ measured under operating mode 1 (V/F–10 Hz) and mode 4 (SLV–10 Hz).

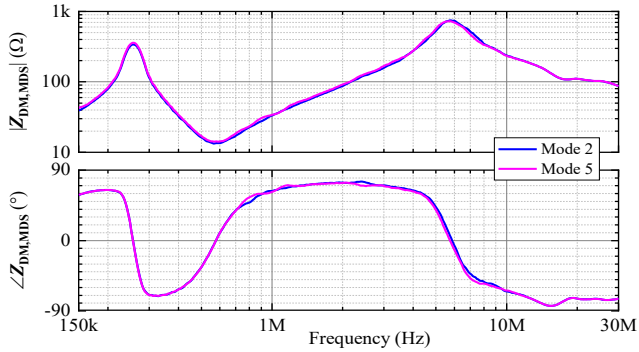

Fig. 8. $Z_{DM,MDS}$ measured under operating mode 2 (V/F–30 Hz) and mode 5 (SLV–30 Hz).

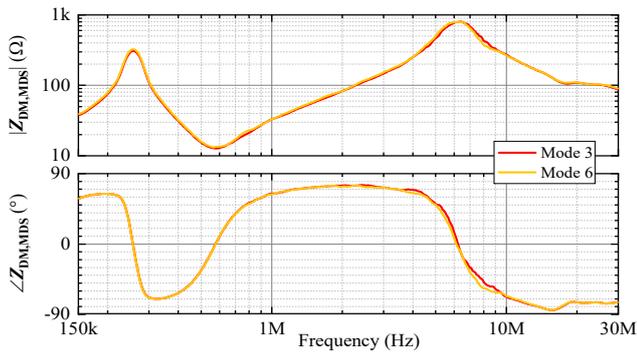

Fig. 9. $Z_{DM,MDS}$ measured under operating mode 3 (V/F–50 Hz) and mode 6 (SLV–50 Hz).

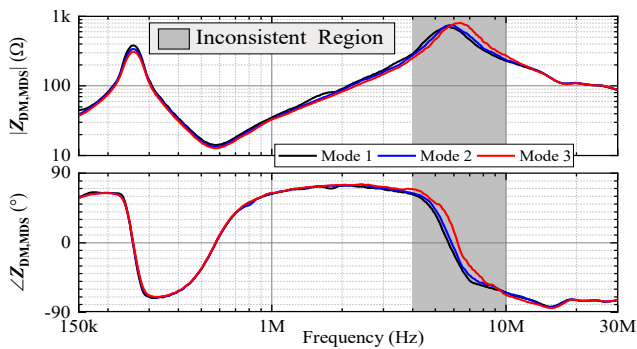

Fig. 10. $Z_{DM,MDS}$ measured under operating mode 1 (V/F–10 Hz), mode 2 (V/F–30 Hz), and mode 3 (V/F–50 Hz).

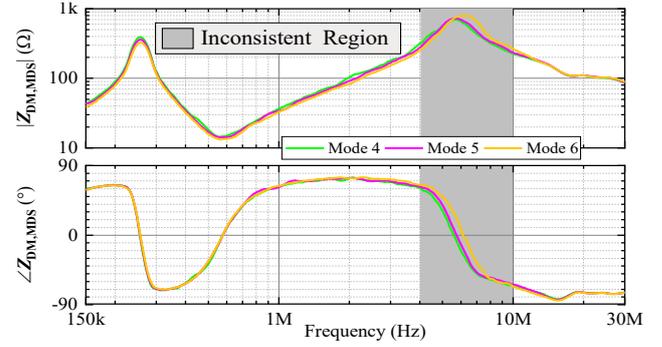

Fig. 11. $Z_{DM,MDS}$ measured under operating mode 4(SLV–10 Hz), mode 5(SLV–30 Hz), and mode 6(SLV–50 Hz).